\documentstyle[aps,prb]{revtex}

\begin{document}

\title{Wetting transitions of $^4$He on alkali metal surfaces from density
functional calculations}
\author{Francesco Ancilotto, Fabio Faccin and Flavio Toigo}
\address{INFM and Dipartimento di Fisica "G. Galilei", via Marzolo 8,\\
I-35131 Padova, Italy}
\maketitle

\bigskip

\begin{abstract}
We have studied the wetting properties of $^4$He adsorbed on the surface of
heavy alkali metals by using a non-local free-energy density-functional
which describes accurately the surface properties of liquid $^4$He in the
temperature range $0<T<3\,K $. Our results for $^4$He on the Cs surface give
both the temperature dependence of the contact angle and the wetting
temperature in good agreement with the experimental findings. For the $^4$%
He/Rb system we find that a wetting transition on the Rb surface occurs at $%
T\sim 1.4 \,K$, whereas the experiments show either wetting down to T=0 or a
wetting transition at $T\sim 0.3\,K$. We suggest that this disagreement is
due either to an inaccuracy of the fluid-substrate potential used in our
calculations or the consequence of substrate roughness, which is known to
affect the Rb surface, and whose effect would be to lower the wetting
temperature. The sensitivity of the wetting transition to the He-surface
potential are stressed for the He/Rb surface, which may justify the
controversial experimental findings for this system.
\end{abstract}

\bigskip
\bigskip

\newpage

\section{Introduction}

The existence and nature of the wetting transition \cite{cahn77} at a
temperature above the triple point $T_{t}$ (and below the critical
temperature) is known to be the result of a delicate balance between the
interatomic potential acting among the atoms in the liquid and the
atom-substrate interaction potential \cite{dietrich}. When the latter
dominates, the fluid tends to wet the surface. Because of the extremely weak
interaction between He atoms, liquid $^{4}$He in contact with almost any
substrate spreads to form a continuous film over the surface, so that vapor
and substrate are never in contact, thus making $^{4}$He the ideal wetting
agent.

A remarkable exception to this behavior is found when $^{4}$He is adsorbed
on heavy alkali metal surfaces: due to the large electron spill-out at their
surface, alkali metals provide the weakest adsorption potentials in nature
for He atoms. Based on this observation, it was suggested \cite{cheng2,saam}
that wetting and prewetting transitions of $^{4}$He ought to occurr on Cs
(which exerts the weakest attraction to He atoms) and, perhaps, Rb
substrates. Actually, experimental evidence that He$^{4}$ on a Cs substrate
exhibits a wetting transition \cite{rutledge,nacher,ketola,klier} at $%
T_{w}\approx 2$ K has been collected. The results for $^{4}$He on Rb are
more controversial, since both a prewetting transition accompanied by
complete wetting down to $T=0$ K \cite{phillips}, (this is analogous to
triple point wetting) and a wetting transition \cite{wyatt} at a nonzero $%
T_{w}\approx 0.3$ K have been reported. On the more attractive substrate of
potassium, wetting behavior has been observed for all $T>0$ \cite{phil}.

Numerical simulations have proven to be a very useful tool to study the
wetting properties of fluids. In particular, Density Functional (DF) methods 
\cite{evans} have become increasingly popular in recent years because of
their ability to describe inhomogeneous fluids and phase equilibria. A
comparison with ''exact'' Monte Carlo simulation results shows that, once
the long-range (van der Waals) attractive forces exerted by a surface on the
gas atoms are included in the free energy functional, Density Functional
methods provide a qualitatively (and most often quantitatively) good
description of the thermodynamics of gas adsorption on a solid surface and
correctly predict a large variety of phase transitions (wetting, prewetting,
layering,etc.).

In particular, a quite accurate description of the $T=0$ properties of
liquid $^{4}$He has been obtained within a DF approach by using the energy
functional proposed in Ref.\cite{dupont} and later improved in Ref.\cite
{prica}. This phenomenological functional has been widely tested in a
variety of problems involving inhomogeneous phases of liquid $^{4}$He
(surfaces, droplets, films, etc.) \cite{saam,prica,dalfovo1,barranco1,anci1}%
, and is known to give a good description of the $T=0$ properties of the
liquid-vapor interface and also of some properties of liquid $^{4}$He on
various substrates \cite{anci,sartori} (for a thorough comparison between
these and other functionals used to describe liquid $^{4}$He at $T=0$, see
Ref. \cite{szybis}). A very remarkable example of success of such
theoretical framework is the prediction of the
non-wetting behavior of liquid $^{4}$He on heavy-alkali metals 
at zero absolute temperature \cite{saam}.

In the present work we extend the theory of Refs.\cite{dupont,prica} to
finite temperatures and apply the functional to the study of wetting
properties of $^{4}$He adsorbed on alkali metal surfaces. A preliminary
account of this work has been presented elsewhere\cite{nist}.

A previous attempt to construct a semi-empirical finite-temperature Density
Functional for $^{4}$He has been reported in Ref.\cite{barranco}. There, a 
{\it local} functional of the density was used (i.e. where the He-He
interaction terms correspond to an attractive two-body contact force), with
a number of parameters adjusted to guarantee the correct liquid-vapor
coexistence. The agreement with the experimental surface tension of liquid $%
^{4}$He at saturated vapor pressure was also imposed by construction. This
functional has been applied to the study of the behavior of the liquid-vapor
interface thickness as a function of temperature \cite{barranco}.

\ It is well known \cite{evans}, however, that a local theory gives incorrect
fluid structures in situations where the liquid is perturbed on a
microscopic scale, as it happens close to a solid substrate. This failure is
due to the zero-range nature of the Skyrme interaction used in Ref.\cite
{barranco}, which completely neglects the short-range correlations
associated with the interatomic potential hard-core.

For this reason, in our work we rather use as a starting point for our
finite-temperature calculations the {\it non-local} form of the Density
Functional proposed in Ref.\cite{prica}, where a more realistic finite-range
interactions between He atoms is included. Contrary to what happens with the
local functional \cite{barranco}, this non-local functional with parameters
fitted to bulk properties provides the correct asymptotic behavior of the
surface potential profile without any ad-hoc prescription, which is a
desirable feature in order to properly describe the liquid-vapor interface.

\section{Method}

In the Density-Functional approach the grand-canonical free-energy is
considered a functional of the density $\rho ({\bf {r})}$:

\begin{equation}
\Omega _{He}[\rho ]=F[\rho ]-\mu \int d{\bf {r\;}}\rho ({\bf {r})}
\label{eq:omega}
\end{equation}
where $\mu $ is the chemical potential and $F[\rho ]$ is the free-energy
density functional.

We propose to write $F[\rho ]$ as: 
\begin{equation}
F=\int f \, d{\bf r}  \label{eq:F1}
\end{equation}

where

\begin{equation}
f\equiv f_{quant}[\rho ]+f_{int}[\rho ]  \label{eq:F}
\end{equation}

The first term in the r.h.s. takes into account the quantum kinetic energy
and the free energy density of an ideal Bose gas $f_{id}$ \cite{huang}:

\begin{equation}
f_{quant}={\frac{\hbar ^{2}}{2M}}(\nabla \sqrt{\rho })^{2}+c_{4}f_{id}
\label{eq:free}
\end{equation}
with $c_{4}$ an adjustable parameters to be determined by the fitting
procedure detailed below.

As in the original T=0 functional \cite{prica}, $f_{int}$ accounts for He-He
interactions (in a Hartree-like fashion) and for correlations effects:

\begin{eqnarray}
f_{int}[\rho ({\bf r})]= &&\frac{1}{2} \int d{\bf r}^{\prime}\rho ({\bf r})
\rho ({\bf r}^{\prime})V_{\ell}(|{\bf r}-{\bf r}^{\prime}|)  \nonumber \\
+&&\frac{c_2}{2}\rho ({\bf r})(\bar{\rho}_{{\bf r}})^{2} +\frac{c_3}{3}\rho (%
{\bf r})(\bar{\rho}_{{\bf r}})^{3}  \nonumber \\
-&& \frac{\hbar ^2}{4M}\alpha _s \int d{\bf r}^{\prime}G(|{\bf r}-{\bf r}%
^{\prime}|) (1-{\frac{\rho ({\bf r})}{\rho _{0s}}})\nabla \rho ({\bf r})
\nabla \rho ({\bf r}^{\prime}) (1-{\frac{\rho ({\bf r}^{\prime})}{\rho _{0s}}%
}).  \label{eq:dft}
\end{eqnarray}

The first term contains a two-body He-He pair potential $V_{l}(r)$ screened
at distances shorter than a characteristic length $h_l$ ($V_{l}=0$ for $%
r<h_l $), while the second and third terms contain $\bar{\rho}_{r}$, i.e.
the average of the density over a sphere of radius $h_l$, and account for
the increasing contribution of the hard-core He-He repulsion when the
density is increased. The last term contains the gradient of the density at
different points and corresponds to a non-local correction to the kinetic
energy.

We remark at this point that the quantum kinetic energy term appearing in (%
\ref{eq:free}) (as well as in other phenomenological density functional
theories used to describe inhomogeneous $^{4}$He) is derived from the {\it %
total} density $\rho ({\bf {r})}$. However, as discussed in Ref.\cite
{griffin}, a complete Density Functional theory of Bose-condensed liquids
would rather involve functionals of both the local density $\rho ({\bf {r})}$
and the local order parameter $\Phi ({\bf {r})=\sqrt{\rho _{c}({r})}}$, $%
\rho _{c}({\bf {r})}$ being the local condensate density. In particular, the
counterpart of the first term in (\ref{eq:free}) would be proportional to $%
\left( \nabla \sqrt{\rho _{c}}\right) ^{2}$. A rationalization of why the
simplified form of the density functional used here (and in all other
applications of \ DF theory to liquid $^{4}$He as well) leads to a
satisfactory description of the surface can be traced to the fact that the
two quantum kinetic energy terms coincide in the low density surface region,
which is important for the interface problems, where $\rho _{c}({\bf {r}%
)\sim \rho ({r})}$ \cite{griffin}.

In the T=0 functional \cite{prica}, the free parameters $h_{l}$, $c_{2}$ and 
$c_{3}$ were adjusted in order to reproduce the experimental values of the
density, of the energy per atom, and of the compressibility for bulk liquid $%
^{4}$He at zero pressure, while the width $l_{G}$ of the gaussian function $G
$ and the parameters $\alpha _{s}$ and $\rho _{0s}$ in Eq. (\ref{eq:dft})
were adjusted to reproduce the overall shape of the experimental static
response function of bulk liquid $^{4}$He. For a detailed description of the
various terms in (\ref{eq:dft}) we refer the reader to Ref. \cite{prica}.

We follow the same philosophy at finite temperatures, i.e. we consider the
free energy density of a uniform system ($\rho $=costant) as given in Eqs. (%
\ref{eq:free},\ref{eq:dft}) :

\begin{equation}
f\equiv F/V={\frac{1}{2}}b\rho ^{2}+{\frac{1}{2}}c_{2}\rho ^{3}+{\frac{1}{3}}%
c_{3}\rho ^{4}+c_{4}f_{id}(\rho ,T)  \label{eq:fhomo}
\end{equation}
where $b=\int d{\bf {r\;}}V_{l}({\bf r}){\bf =}4\pi \int_{h_{l}}^{\infty }d%
r\;r^{2}V_{l}({\bf r}){\bf ,}$ and minimize the grand potential with
respect to $\rho $ to get the chemical potential:

\begin{equation}
\mu (\rho ,T)\equiv {\frac{\partial f}{\partial \rho }}=b\rho +{\frac{3}{2}}%
c_{2}\rho ^{2}+{\frac{4}{3}}c_{3}\rho ^{3}+c_{4}\mu _{id}  \label{eq:mu}
\end{equation}

and the pressure

\begin{equation}
P(\rho ,T)\equiv \rho \mu -f={\frac{1}{2}}b\rho ^{2}+c_{2}\rho
^{3}+c_{3}\rho ^{4}+c_{4}P_{id}  \label{eq:press}
\end{equation}

If $\rho _{v}$ and $\rho _{l}$ are the densities of vapor and liquid \ $^{4}$%
He at saturated vapor pressure at a given temperature T, than at coexistence
the equilibrium conditions:

\begin{equation}
\mu (\rho _{l},T)=\mu (\rho _{v},T)  \label{eq:mu12}
\end{equation}
and

\begin{equation}
P(\rho _{l},T)=P(\rho _{v},T)  \label{eq:pexp}
\end{equation}
must be satisfied.

By requiring that equalities (\ref{eq:mu12}),(\ref{eq:pexp}) are satisfied
when the experimental values of $\rho _{v}$ and $\rho _{l}$ are inserted
into Eqs. (\ref{eq:mu}),(\ref{eq:press}) and, moreover, that the common
value of the chemical potentials in Eq. (\ref{eq:mu12}) \ is equal to the
experimental value at the same temperature\cite{maynard}:

\begin{equation}
\mu (\rho _{l},T)=\mu ^{expt}(T),  \label{eq:muexp}
\end{equation}
we get three equations relating the four adjustable coefficients $c_{2}$, $%
c_{3}$, $c_{4}$, and $b.$

As a fourth equation necessary to determine them completely at any
temperature of interest we impose that the calculated isothermal
compressibility, $1/K\equiv \rho (\partial P/\partial \rho )$, reproduces
its corresponding experimental value:

\begin{equation}
K(\rho _{l},T)=K^{expt}(T)  \label{eq:compr}
\end{equation}

The fit has been carried out in the region $0<T<3\,K$, which is the most
interesting for the wetting properties of $^{4}$He we want to investigate.
Once $b(T)$ has been obtained by solving the system of four equations
described above, a straightforward calculations provides $h_{l}(T)$ so that
all coefficients entering the functional (\ref{eq:dft}) are determined. The
calculated coefficients are shown, for some selected temperatures, in Table (%
\ref{table:param}).

We wish to stress at this point that all unknown coefficients entering the
free energy Density Functional of pure $^{4}$He are fitted to bulk
properties, i.e. to properties of uniform systems.

To include the interaction with the substrate, we use a binding potential $%
V_{s}(z)$ which describes the interaction between the alkali metal,
occupying the half space $z\leq 0$, and one $^{4}$He atom located at a
distance $z$ above the ideally flat surface.

The physisorption potential $V_{s}(z)$ is taken in the form originally
proposed in Ref.\cite{kohn}, i.e. as a sum of a Hartree-Fock repulsion and a
van der Waals attraction. Both terms are parametrized with a set of
coefficients. Very reliable values for the parameters entering $V_{s}(z)$
and describing the interaction of $^{4}$He with alkali substrates have been
calculated from first principles \cite{chizmeshya}. These potential have
been used to calculate the wetting properties of rare gases (other than He)
on alkali metal surfaces \cite{anciprew} and also the T=0 profile of a
droplet on the Cs surface \cite{sartori}, and found to give predictions in
reasonable agreement with experiments for all these systems.

The total free energy functional for $^4$He interacting with the substrate
is thus:

\begin{equation}
\Omega [\rho ({\bf r})]=\Omega _{He}[\rho ({\bf r})] + \int d{\bf r}\rho (%
{\bf r}) V_s(z).  \label{eq:etot}
\end{equation}

According to Density Functional theory, the equilibrium density profile $%
\rho ({\bf {r})}$ of the fluid in the presence of the substrate can be
determined by applying the variational principle :

\begin{equation}
\delta \Omega /\delta \rho ({\bf r})=0  \label{eq:delta}
\end{equation}
to get the Euler-Lagrangian equation:

\begin{equation}
\{-{\frac{\hbar ^{2}}{2M}}\nabla ^{2}+U[\rho ({\bf r})]+V_{s}(z)\}\sqrt{\rho
({\bf r})}=\mu \sqrt{\rho ({\bf r})},  \label{eq:kseq}
\end{equation}
where the effective potential $U$ is defined as $U[\rho ({\bf r})]\equiv
\delta F[\rho ({\bf r})]/\delta \sqrt{\rho }\,\ $and the value of the
chemical potential $\mu $ is fixed by the normalization condition

\begin{equation}
\int d{\bf r}\rho ({\bf r})=N,  \label{eq:norm}
\end{equation}
$N$ \ being the total number of $^{4}$He atoms.

\section{Results and Discussion}

Wetting or non wetting is determined by surface tension balancing. The
intersection of a macroscopic liquid droplet with a solid planar substrate
can be characterized by the contact angle $\Theta $: it goes to zero as $T$
approaches $T_{w}$ and remains zero at all higher temperatures, where the
liquid wets the substrate. Below the wetting temperature, the contact angle
is determined by balancing the forces acting along the contact line and
depends on the interfacial tensions $\sigma _{ij}$ between each pair of
coexisting phases through  Young-Dupre's equation

\begin{equation}
cos\Theta ={\frac{\sigma _{sv} - \sigma _{sl} }{\sigma _{lv} }}.
\label{eq:young}
\end{equation}
The subscripts $l$, $v$ and $s$ identify the liquid, vapor and solid,
respectively.

We have calculated the interfacial tensions $\sigma _{ij}$ entering Eq. (\ref
{eq:young}) at selected values of $T$ by using the density functional method
described in the previous section.

We first studied the liquid-vapor interface of $^{4}$He by solving Eq. (\ref
{eq:kseq}) (with $V_{s}\equiv 0$) with the appropriate boundary conditions,
corresponding to a planar liquid film in equilibrium with its own vapor. In
this case the solution depends only on the coordinate $z$ normal to the
surface, and provides the equilibrium density profile $\rho (z)$. Our
calculations are performed within a region of length $z_{m}$. Rather large
values of $z_{m}$ and of the film thickness $t_{f}$ are necessary for well
converged calculations (typically $z_{m}\sim 200\,\AA $ and $t_{f}\sim 150\,%
\AA $), in order to let the system spontaneously reach (i) the liquid bulk
density in the interior of the film and (ii) the vapor bulk density in the
region outside the film. In Fig.1 we show some selected density profiles
close to the liquid-vapor interface. A detailed analysis of the resulting
density profiles shows that the width of the interface increases
monotonically with temperature. A comparison of our results with those
obtained in Ref. \cite{barranco} indicates also that the thickness of the
liquid-vapor interface, at a given temperature, is sligthly smaller than
that obtained from the zero-range functional of Ref.\cite{barranco}.

From the calculated equilibrium density profiles one can directly compute
the liquid-vapor surface tension $\sigma _{lv}$, from the definition $\sigma
=(\Omega +PV)/A$. Here $P$ is the saturated vapor pressure at temperature T, 
$V$ is the volume of the system and $A$ is the surface area. For the
one-dimensional problem considered here, one can write:

\begin{equation}
\sigma = (1/2) [\int _0 ^{z_m} f(\rho(z))dz - \mu \int _0 ^{z_m} \rho(z) dz
+Pz_m]  \label{eq:sigma_1d}
\end{equation}

The pressure $P$ can be conveniently calculated from the definition $P=(\mu
\rho _{v}-f(\rho _{v}))=(\mu \rho _{l}-f(\rho _{l}))$, $\rho _{v}$ and $\rho
_{l}$ being the experimental densities for  bulk liquid and vapor,
respectively. The chemical potential $\mu $ is an output of our calculation,
being fixed by the areal density $n_{c}\equiv N/A=\int_{0}^{z_{m}}\rho (z)dz$%
. Finally, a factor 1/2 appears in the previous equation to account for the
two free surfaces delimiting the liquid film in our "slab" calculations.

The dependence of the liquid-vapor surface tension of $\ ^{4}$He on
temperature has been extensively investigated in the past. Surprisingly, the
absolute value of $\sigma _{lv}$ at $T=0$ remains poorly known: predictions
from different groups give values for $\sigma _{lv}(0)$ differing by up to $%
6\%$ \cite{surf1,surf2,surf3,surf4}. Our calculated values are shown in
Fig.2, where they are compared with the available experimental data. The
overall agreement is quite satisfactory, given the fact that our functional
has been fitted to reproduce $^{4}$He bulk properties only. Note also that
the kink (barely) visible in the experimental data at the temperature
corresponding to the $\lambda $-point $T_{\lambda }=2.17\,K$ appears also in
our calculated $\sigma _{lv}$ and reflects similar kinks appearing close to $%
T_{\lambda }$ in the bulk quantities (density, chemical potential and
compressibility) used to fit the coefficients entering the free energy (\ref
{eq:F}).

We next turn to the problem of liquid $^{4}$He in the presence of a solid
surface. The two additional surface tensions $\sigma _{sv}$ and $\sigma _{sl}
$ required to obtain the contact angle $\Theta $ from Eq. (\ref{eq:young})
are calculated within our density functional approximation by using
different boundary conditions from those used to calculate $\sigma _{lv}$.
We  confine the fluid between two parallel surfaces, separated by a distance
large enough to avoid any compression effect. The two surfaces act on the
fluid with the external potential $V_{s}(z)$ appropriate to the alkali metal
substrate. For selected values of the temperature $T$, we get the structure
of the solid-vapor and solid-liquid interfaces by minimizing the
grand-potential (\ref{eq:etot}) subject to the additional constraint that
the density in a large region between the two surfaces (where bulk behavior
is expected) is equal respectively to $\rho _{v}$ and $\rho _{l}$. The
corresponding surface excess\ energies per unit area $\sigma =(\Omega
+PV)/(2A)$ are the solid-vapor $\sigma _{sv}$ and the solid-liquid $\sigma
_{sl}$ surface tensions. Of course the distance between the two confining
surfaces must be such that the calculated values of \ $\sigma _{sv}$ and  
$\sigma _{sl}$ do not change by further increasing their separation. We show
in Fig. 3 the equilibrium density profiles of $^{4}$He confined between two
Rb surfaces, at $T=1.4\,K$. The solid line shows the density for the liquid
phase confined between the surfaces, the dotted line (which is rescaled for
clarity by a factor $30$) shows instead the vapor density profile. Note the
oscillations of the liquid density close to the surface, due to
close-packing effects. We have checked, by means of similar calculations
with different cell sizes, that at least $150\,\AA $ must separate the two
attractive surfaces so that the two fluid-solid interfaces are effectively
decoupled and the calculation gives a converged value for the surface
tensions $\sigma _{sl}$ and $\sigma _{sv}$.

According to Eq. (\ref{eq:young}), wetting of a surface occurs at a
temperature $T_{w}$ such that

\begin{equation}
\sigma _{sv}-\sigma _{sl} = \sigma _{lv}  \label{eq:wett}
\end{equation}

In Fig. 4 we show the behavior of the two sides of Eq.(\ref{eq:wett}) as a
function of temperature, for three different alkali surfaces. A common
feature of the curves for $\sigma _{sv}-\sigma _{sl}$ shown in Fig. 4 is
their rather weak temperature dependence, as indeed observed in experiments 
\cite{rolley}. The predicted wetting temperature for Cs, $T_{w}\sim 2.1\,K$,
is in good agreement with the experimental value $T_{w}\sim 2\,K$. Also in
the case of $^{4}$He on K our calculated results shown in Fig.4 agree with
experiments \cite{phillips}, indicating wetting at any $T>0$. On the
contrary, for Rb our calculations predict a wetting temperature $T_{w}\sim
1.4\,$\ K, in marked disagreement with the experimental findings which show
instead either complete wetting down to $T=0$ \cite{phillips}, or a wetting
transition at $T_{w}\approx 0.3$ K \cite{wyatt}.

We show in Fig.\ 5 the temperature dependence of the contact angle for the $%
^{4}$He/Cs system as obtained from our calculations, together with the
available experimental data. Although the equilibrium contact angle is a
well defined thermodynamic quantity, measurements of this quantity
invariably show a strong hysteretic behavior, i.e. the measured value of the
contact angle depends on whether the contact line is advancing or receding.
For relatively homogenous surfaces, however, it has been shown that the
advancing contact angle is a good measure of the equilibrium contact angle,
so we compare our results with the experimental values for that quantity.
The contact angle for $^{4}$He on Cs surfaces has been measured by different
groups in recent years by using different techniques \cite
{klier,rolley,science,ross1}. In Ref.\cite{science,ross1} the contact angle
is measured by direct visual inspection of macroscopic droplets of
superfluid $^{4}$He on a Cs surface. In Ref.\cite{rolley} the contact angle
is obtained by means of direct optical measurement using interferometric
techniques \cite{rolley}, whereas in Ref. \cite{klier} it is measured, much
more indirectly, from the reduction in pressure due to capillarity, when an
array of parallel tungsten plates coated with Cesium are immersed into
liquid $^{4}$He. Our results are compared in Fig.5 with the aforementioned
experimental data. The overall agreement with the results from Ref.\cite
{rolley,science,ross1} is satisfying, although a strong discrepancy exist
with the data of Ref.\cite{klier}.

At variance with the overall satisfactory quantitative agreement  we get for
the $^{4}$He/Cs system, our predictions seems to be wrong for the $^{4}$%
He/Rb system. We believe that this disagreement is probably due to the fact
that our assumptions of a rigid and planar substrate are far from being
satisfied in this case. Impurities or roughness, in fact, are always present
on the surface, as indicated by the strong hysterectic behaviour of the
contact angle observed in all experiments when studying advancing or
receding contact lines. This condition may be particularly severe in the
case of the Rb surface, where defects and roughness are known to be present
to a larger extent than on the surface of Cs.

The heterogeneity of the substrate is known to promote wetting, i.e. leads
to values of $\sigma _{sv}-\sigma _{sl}$ higher than for an ideal substrate.
In the case of $^4$He on Rb, as apparent from Fig.4, an increase of $\sigma
_{sv}-\sigma _{sl}$ would shift the wetting temperature towards lower
values, as indeed found in the experiments.

The characterization of the wetting behavior of superfluid Helium on rough
surfaces is a very difficult task. The quality of the substrates used in the
experiments cannot be determined at lengthscales smaller than optical
wavelengths. The theoretical description of the effects of heterogeneities
is also challenging. Only few theoretical results are available on the
effect of microscopic disorder \cite{curta,tang} on the wetting properties
of fluids adsorbed on weakly attractive surfaces.

The effect of model disorder (on an atomic scale) on the wetting properties
of  Ne on the Mg surface has been studied by Curtarolo et al. \cite{curta}.
One interesting conclusion contained in this paper is that roughness destroy
the discontinuous nature of the wetting transition, even if a trace of
discontinuities remain in the calculated isotherms. 

In Ref.\cite{tang} a Lennard-Jones fluid and its wetting properties on
molecularly rough surfaces have been studied by using Molecular Dynamics.
Contrary to the conventional belief that surface roughness reduces the
contact angle thus making it easier to wet a rough surface than a smooth
one, the surprising conclusion of this work is that the contact angle is
larger for the rough surface than for the smooth one.  

In order to have further insight into this important issue,
we are currently investigating the effect of a realistic distribution of
microscopic disorder on the wetting transition of $^{4}$He on the Rb surface
within the Density Functional approach described here \cite{gatica}.

An alternative explanation for the disagreement with the experiments for the 
$^4$He/Rb system is that the He-Rb surface potential used in our
calculations is inaccurate. Our overestimate of the wetting temperature
would be a sign, in this case, that the calculated potential is slightly
less attractive than required to provide a result closer to the experimental
one.

We have thus tried a modified He/Rb surface interaction by changing sligtly
the two parameters entering the repulsive part of the ab-initio potential $%
V_{s}(z)$ describing the He-surface interaction. The modified potential, has
a minimum of depth $D=0.661\,$ meV located at $z_{m}in=8.46\,a_{0}$, to be
compared with the values $D=0.629\,$ meV and $z_{m}in=8.60a_{0}$ obtained
with the original parameters. Such a tiny variation in the adsorption
potential ($\sim 5\,\%$ increase in the
potential well depth) has quite a large 
effect on the He/Rb wetting temperature, as can
be judged from the curve labeled $Rb^{\ast }$ in Fig. 4. With this modified
potential the wetting transition  moves very close to $T=0$, as in the
experiments. Interestingly, in this case the curve $-\sigma _{sl}(T)$
almost coincides with the curve $\sigma _{lv}(T)$ for a range of $T$
values, between $0$ and $0.3-0.4$ K (at such temperatures the contribution $%
\sigma _{sv}$ is negligible in our approximations). This could provide
a reason for the different experimental estimates of the wetting temperature
(either $T=0$ or $T\sim 0.3-0.4\,\ $K) by different groups: in the region
where the two curves almost overlap their slope is very small, and thus
small variations in the quality of the substrate may result in rather
different wetting temperatures. Moreover, this could also suggest a
different character of the wetting transition for He/Rb,
since similar slopes in $-\sigma _{sl}(T)$ and $\sigma _{lv}(T)$ imply
a small discontinuity in $d\, cos(\Theta )/dT$, i.e. a 
quasi-continuous transition.

\section{Summary}

We have extended to finite temperatures the non-local Density Functional
proposed in recent years to describe the $^{4}$He properties at T=0 \cite
{dupont,prica}, and studied the wetting behavior of $^{4}$He adsorbed on
alkali metal surfaces. The Density Functional depends on a number of
temperature-dependent phenomenological parameters which are adjusted to
reproduce {\it bulk} experimental properties of liquid $^{4}$He at saturated
vapor pressure. We find that the resulting functional describes accurately
the properties of the liquid-vapor interface of the free $^{4}$He surface.
In the presence of an external potential simulating a planar, solid surface
whose binding properties reproduce those of alkali metal surfaces, it also
accurately reproduces the properties of the $^{4}$He/Cs system, giving the
temperature dependence of the contact angle in good agreement with
experiments. The contact angle vanishes at $T\sim 2.1\,K$, in agreement with
the experimental measure for the wetting temperature, $T\sim 2K$. For $^{4}$%
He on a K surface, we find wetting at all nonzero temperatures, as found
also in experiments. For $^{4}$He on the Rb surface, however, our
calculations fail to reproduce quantitatively the experiments, indicating a
wetting transition at $T\sim 1.4K$. The discrepancy with experiments could
be  either due to our neglect of surface inhomogeneities, which are known to
be present on this surface to a larger extent than on a Cs surface, or  to
some inaccuracy in the theoretical determination of the surface-adatom
potential used in our calculations. If this is the case, our results
indicate that the correct potential should be slightly more attractive than
currently believed.

\bigskip

Acknowledgments:

We thank M.~Barranco, M.W.~Cole, G.~Mistura and L.~Bruschi for useful
comments and discussions.

\newpage

\centerline{ \bf Figure captions:} \bigskip

Figure 1: Calculated $^4$He density profile near the liquid-vapor interface.
Solid line: $T=0\,K$; dotted line: $T=1.5\,K$; Dashed line: $T=2\,K$;
Dash-dotted line: $T=2.5\,K$.

\bigskip

Figure 2: Surface tension of liquid $^4$He. Dots: calculated $\sigma _{lv}$;
Squares, diamonds, triangles and crosses show the experimental results from
Ref.\cite{surf1}, \cite{surf2},\cite{surf3} and \cite{surf4}, respectively.
Lines are only a guide to the eye.

\bigskip

Figure 3: Equilibrium density profiles used to calculate $\sigma _{sl}$
(upper curve, with solid line), and $\sigma _{sv}$ (lower curve, with dotted
line), respectively.

\bigskip

Figure 4: Calculated surface tensions for the $^4$He/alkali systems.
Squares: $\sigma _{lv}$; Dots, diamonds and crosses show ($\sigma
_{sv}-\sigma _{sl}$) for $^4$He/Cs, $^4$He/Rb and $^4$He/K, respectively.
The set of points labeled Rb$^\ast $ are obtained by using a modified $^4$%
He/Rb surface interaction, as explained in the text. Lines are only a guide
to the eye.

Figure 5: Temperature dependence of the $^4$He/Cs contact angle: squares:
this work; triangles, dots and crosses show the experimental results from
Ref.\cite{rolley}, \cite{ross1} and \cite{klier}, respectively.

\newpage

\begin{table}[tbp]
\caption{ }
\label{table:param}
\begin{tabular}{cccccccc}
$T \,( K ) $ & $\rho _l$ $(\AA ^{-3})$ & $\rho _v $ $(\AA ^{-3})$ & $b$ $(K
\AA ^{3})$ & $c_2$ $(K \AA ^{6})$ & $c_3$ $(K \AA ^{9})$ & $c_4$ & $h_l$ $%
(\AA )$ \\ \hline
0.0 & 0.0218354 & 0.0 & -719.2435 & -24258.88 & 1865257 & 0.0 & 2.19035 \\ 
0.4 & 0.0218351 & 0.441987(-11) & -714.2174 & -24566.29 & 1873203 & 0.98004
& 2.18982 \\ 
0.6 & 0.0218346 & 0.452619(-8) & -705.1319 & -25124.17 & 1887707 & 0.99915 & 
2.18887 \\ 
0.8 & 0.0218331 & 0.133640(-6) & -690.4745 & -26027.12 & 1911283 & 0.99548 & 
2.18735 \\ 
1.2 & 0.0218298 & 0.495337(-5) & -646.5135 & -28582.81 & 1973737 & 0.99666 & 
2.18287 \\ 
1.4 & 0.0218332 & 0.147990(-4) & -625.8123 & -29434.03 & 1984068 & 0.99829 & 
2.18080 \\ 
1.6 & 0.0218453 & 0.346234(-4) & -605.9788 & -30025.96 & 1980898 & 1.00087 & 
2.17885 \\ 
1.8 & 0.0218703 & 0.684096(-4) & -593.8289 & -29807.56 & 1945685 & 1.00443 & 
2.17766 \\ 
2.0 & 0.0219153 & 0.119400(-3) & -600.8313 & -27850.96 & 1847407 & 1.00919 & 
2.17834 \\ 
2.1 & 0.0219500 & 0.151732(-3) & -620.9129 & -25418.15 & 1747494 & 1.01156 & 
2.18032 \\ 
2.2 & 0.0219859 & 0.188202(-3) & -619.2016 & -25096.68 & 1720802 & 1.01436 & 
2.18015 \\ 
2.4 & 0.0218748 & 0.275044(-3) & -609.0757 & -26009.98 & 1747943 & 1.02130 & 
2.17915 \\ 
2.6 & 0.0217135 & 0.383550(-3) & -634.0664 & -23790.66 & 1670707 & 1.02770 & 
2.18162 \\ 
2.8 & 0.0215090 & 0.516536(-3) & -663.9942 & -21046.37 & 1574611 & 1.03429 & 
2.18463 \\ 
3.0 & 0.0212593 & 0.676814(-3) & -673.6543 & -20022.76 & 1535887 & 1.04271 & 
2.18562
\end{tabular}
\end{table}

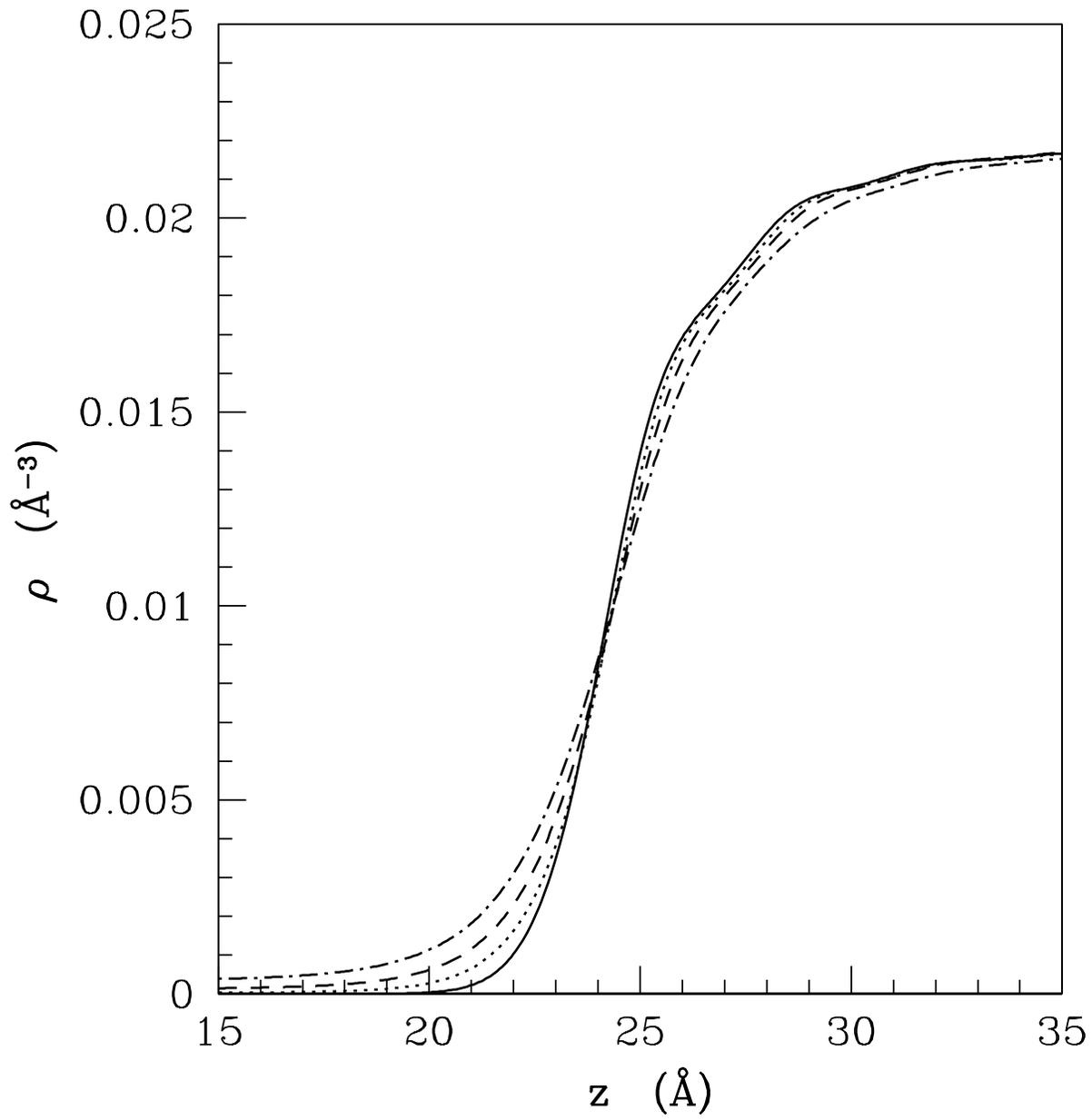
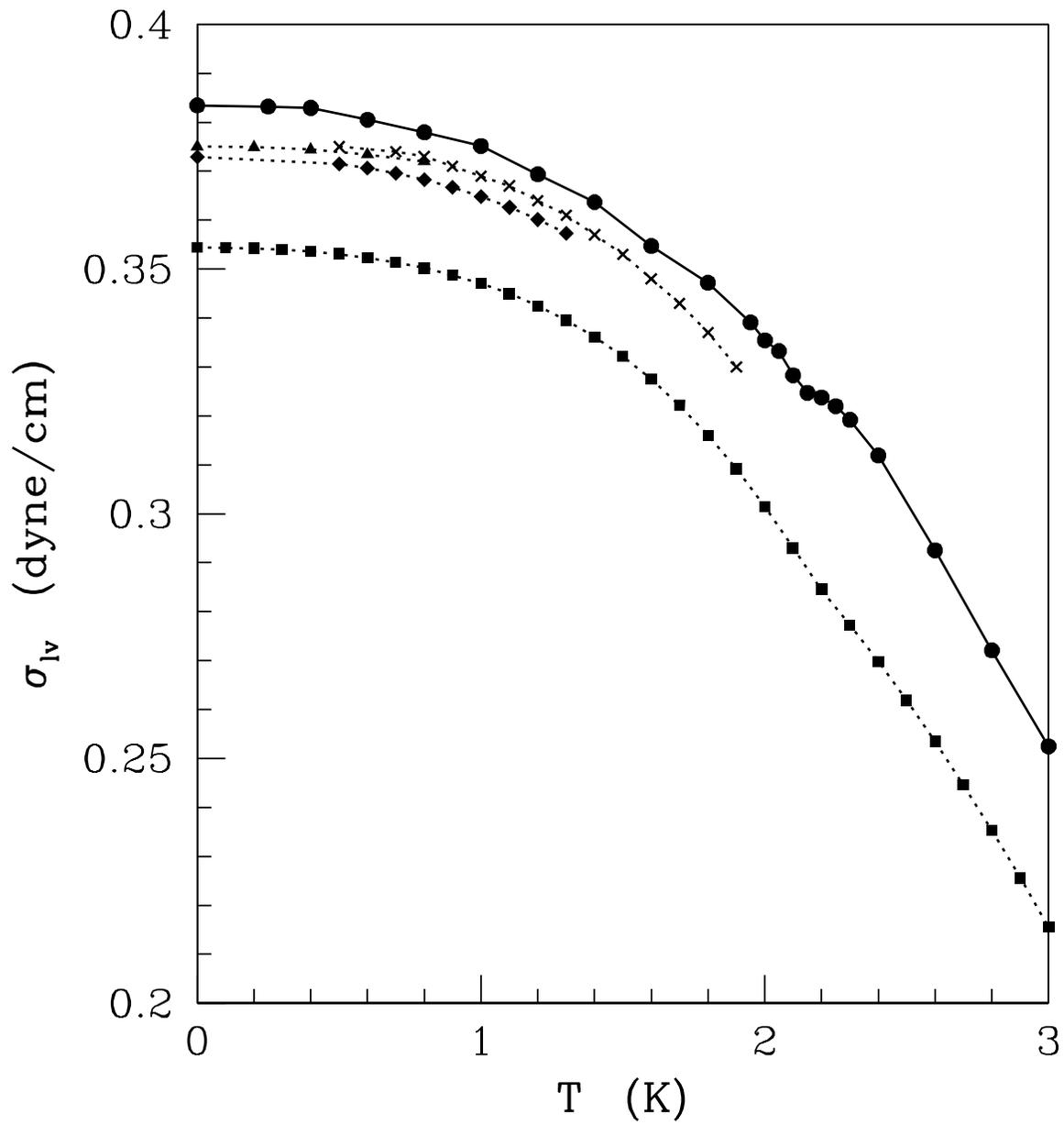
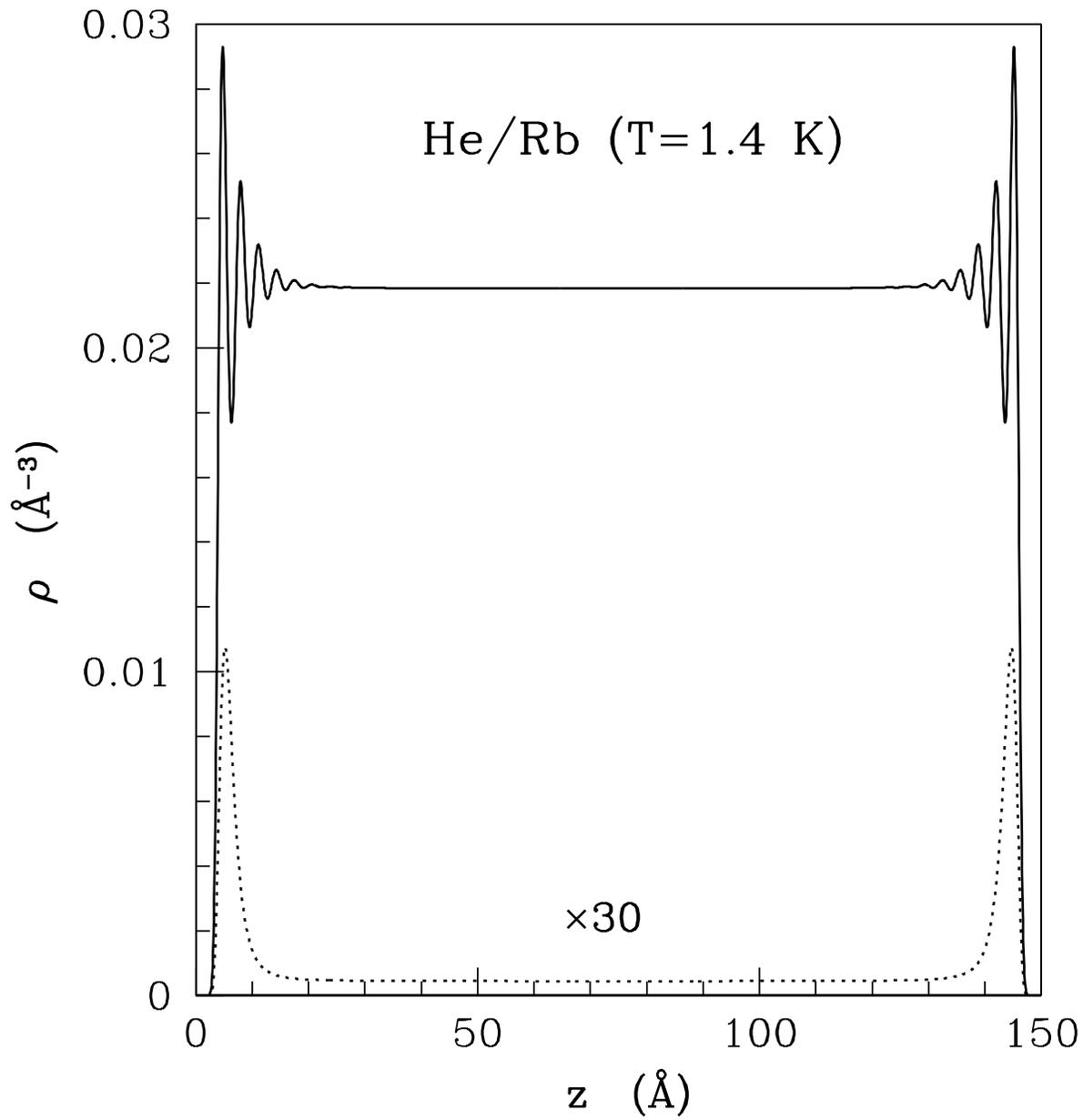
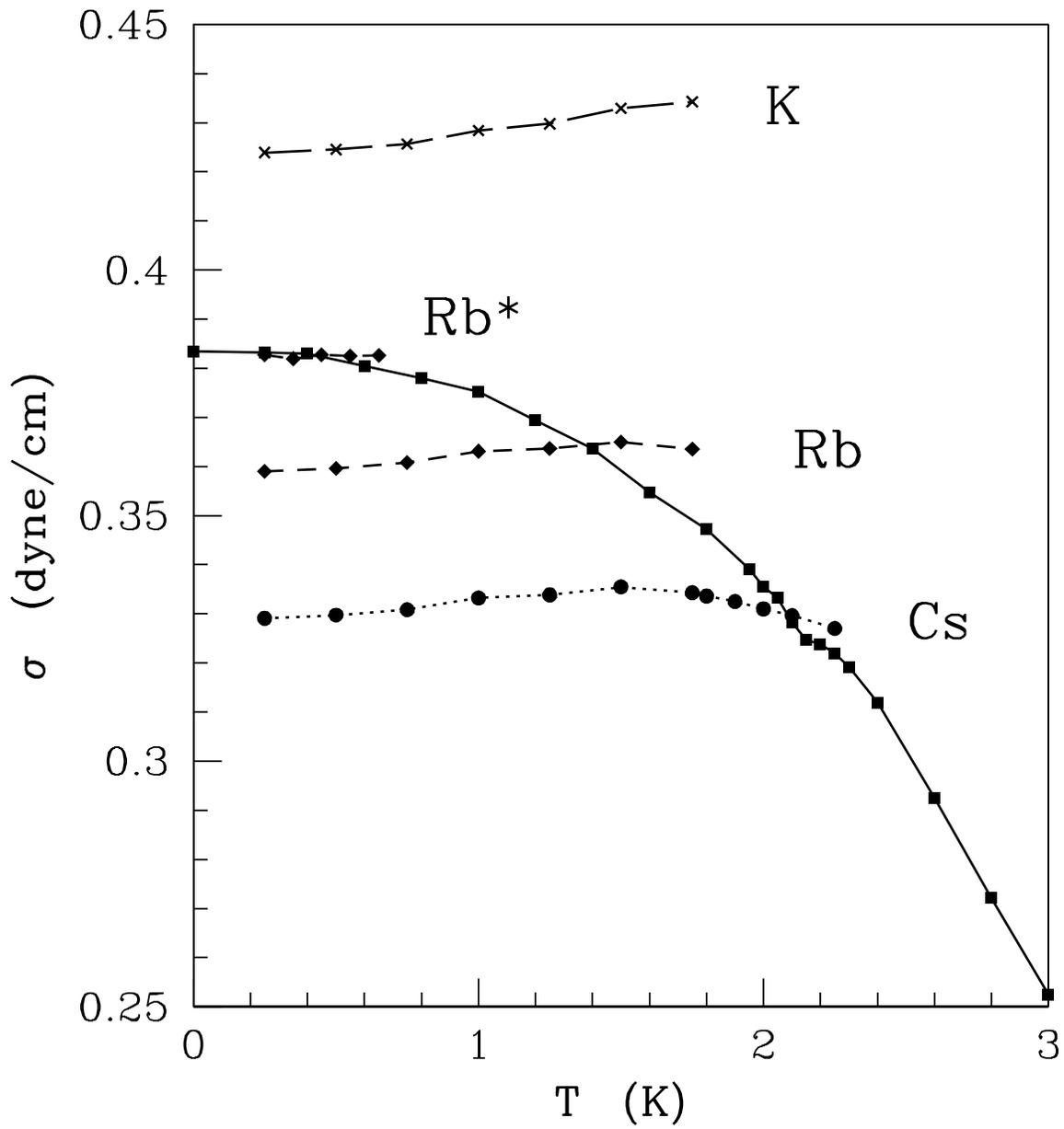
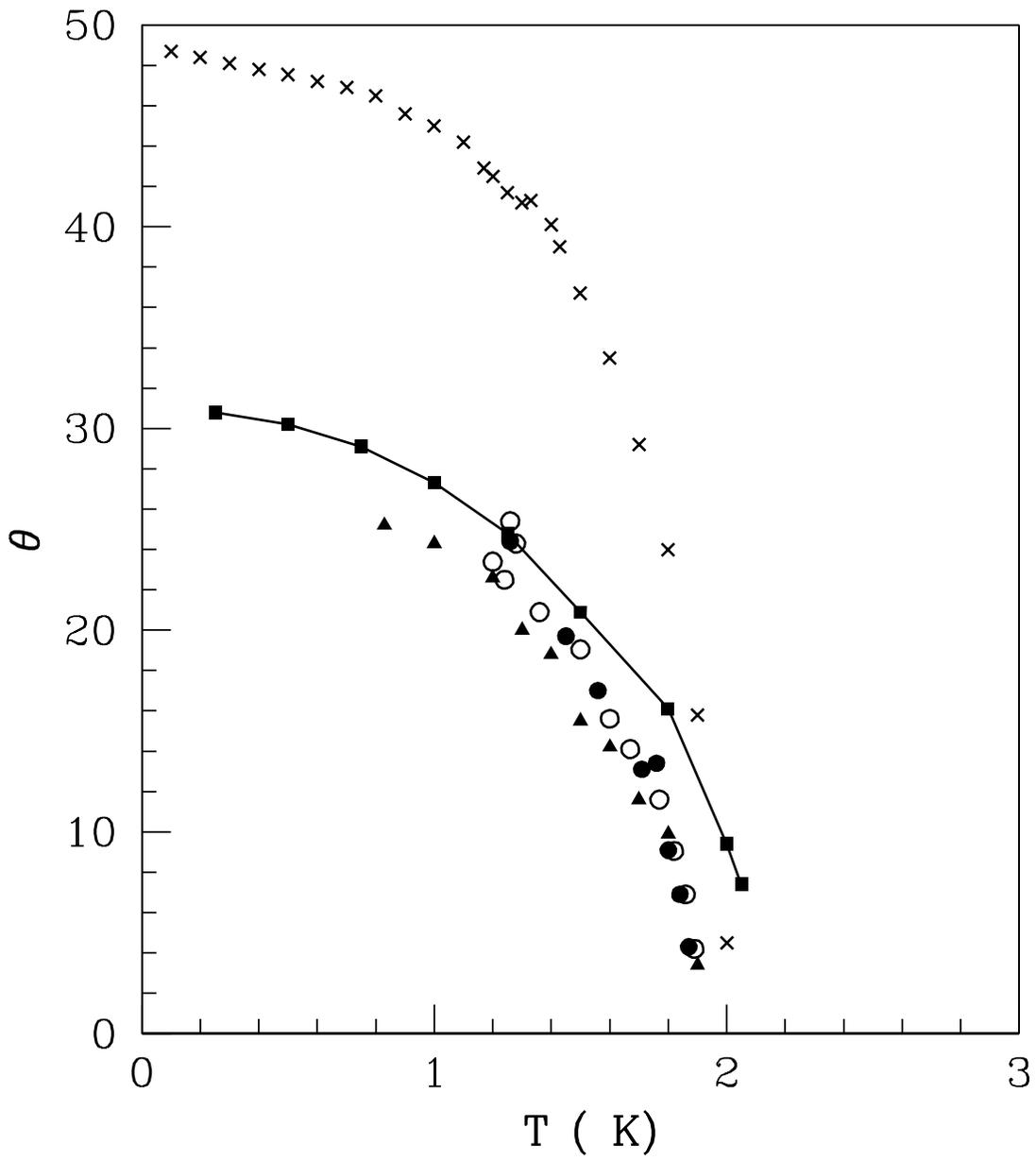
\end{document}